\documentclass[11pt,english]{iopart} 
\usepackage[english]{babel}
\usepackage{amsfonts,amssymb,bm,latexsym}

\usepackage[T1]{fontenc}
\usepackage[latin3]{inputenc}

\newcounter{llista}

\begin{document}
\jl{6}

\title{On the motion of a classical charged particle}

\author{J.\ M.\ Aguirregabiria\dag, J.\ Llosa\ddag  
\ and A.\ Molina\ddag
}
\address{\dag\ Fisika Teorikoa, Zientzia eta Teknologia Fakultatea, Euskal Herriko Unibertsitatea, P.K. 644, E48080 Bilbao (Spain)}
\address{\ddag\ Departament de F\'{\i}sica Fonamental, Universitat de Barcelona\footnote{Postal address: Martí i Franquès, 1; E-08028  Barcelona (Spain)} and Laboratori de F\'{\i}sica Matem\`atica, SCF (Institut d'Estudis Catalans)}
\ead{pitu.llosa@ub.edu}

\begin{abstract}
We show that the Lorentz-Dirac equation is not an unavoidable
consequence of energy-momentum conservation for a point charge. What
follows solely from conservation laws is a less restrictive equation
already obtained by Honig and Szamosi. The latter is not properly an
equation of motion because, as it contains an extra scalar variable, it
does not determine the future evolution of the charge. We show that a
supplementary {\em constitutive relation} can be added so that the
motion is determined and free from the troubles that are customary in
Lorentz-Dirac equation, i.\ e.\ preacceleration and runaways. 
\end{abstract}


\maketitle

\section{Introduction}

Lorentz-Dirac equation is widely accepted as the classical equation of
motion of an elementary point charge interacting with its own radiation
(see for instance \cite{Dirac38,Rohrlich65,Rowe75,Teitel79}):
\begin{equation}
m a^\mu = F^\mu +\frac{2 e^2}{3 c^3}\,\left(\dot a^\mu -\frac1{c^2} \,a^\lambda a_\lambda v^\mu \right),
\label{e1}
\end{equation}
where $F^\mu = \frac{e}{c} F^{\mu\nu}_{\rm ext} v_\nu $ is the external electromagnetic force. 

It is also well known that this equation is affected by some
irreconciliable difficulties, that already show up in the case of
rectilinear motion. Consider a free point charge that enters
perpendicularly a parallel-plate capacitor at $\tau=0$ (proper time) and
leaves it at $\tau_1 >0$. For $\tau <0$ the charge is free, $f^\mu=0$
and the solution to (\ref{e1}) is a uniform rectilinear motion,
$a^\mu=0$. We can therefore take $a^\mu(0)=0$ and $v^\mu(0)=v^\mu_{\rm
in}$ as initial data to integrate equation (\ref{e1}), so obtaining a
unique solution for the velocity $v^\mu$. Nevertheless, this solution
has the drawback that, not only $a^\mu(\tau)$ does not vanish for
$\tau>\tau_1$ (when the external action has ceased), but it grows
exponentially for $\tau\rightarrow \infty$, what is known as {\em
runaway solution}.

Rohrlich \cite{Rohrlich65} put forward a way out consisting in that
(\ref{e1}) is not the equation of motion, but it must be supplemented
with an asymptotic condition: if the external force $f^\mu$
asymptotically vanishes, then the acceleration $a^\mu$ asymptotically
vanishes too. As a result the resulting equation of motion is of
integro-differential type and runaway solutions are ruled out (see also \cite{ibison}).

This alternative however implies what is called {\em preacceleration}.
Although the external force vanishes for $\tau<0$, the solution to the
above integro-differential equation presents non-vanishing acceleration
before the force starts. This is not a surprising feature because, as
pointed out in \cite{Blanco95}, it is a consequence of demanding the
asymptotic condition in the future: the integro-differential equation of
motion itself ``foresees'' what will happen in the future, $\tau >
\tau_1$.

It thus seems as though we were facing the following dilemma
\cite{Blanco95}: either (a) classical electrodynamics is
self-contradictory or (b) Lorentz-Dirac equation is not the right
equation that follows from classical electrodynamics.

In view of this dilemma different stances  are found in the literature.
Rohrlich \cite{Rohrlich65} adopts the alternative (a) and adds that this
is not a major trouble because the time scale at which preacceleration
shows up is too small ($\tau_0\approx 10^{-23}\,$s for electrons) far
beyond the limits of validity of the classical theory. He further
stresses that \cite{Rohrlich97} \guillemotleft the notion of ``classical
point charge'' is an oxymoron \ldots \guillemotright\ since classical
physics ceases to be valid below Compton wavelength.
Moniz and Sharp also argued \cite{Moniz74,Moniz77a,Moniz77b} that classical
electrodynamics is only consistent in describing the motion of charges with radius larger
than the classical electron radius, while the quantum theory 
of \emph{nonrelativistic} charges is free of runaways and preacceleration.

Other authors \cite{Valent88,Yaghjian} embrace the alternative
(b) on the basis that the derivation of Lorentz-Dirac equation involves
Taylor expansions and therefore presumes that both the charge worldline
and the external force  are analytic functions. As a consequence,
equation (\ref{e1}) is not valid in those points where $x^\mu(\tau)$ and
$f^\mu(\tau)$ are not analytic. Particularly, Yaghjian \cite{Yaghjian}
studies a charged spherical shell of radius $\epsilon$ and obtains an
alternative equation:
$$ m a^\mu = f^\mu +\frac{2 e^2}{3 c^3}\,\eta(\tau)\,\left(\dot a^\mu -\frac1{c^2} \,a^\lambda a_\lambda v^\mu \right)
$$
where $\eta(\tau)=0$ for $\tau<0$ and $\eta(\tau)=1$ for $\tau\geq2\epsilon/c$.
In another approach \cite{Kerner,Sanz,Bel,JPA1,JPA2,CPC,Spohn}, the Lorentz-Dirac equation is thought of as a 
necessary ---but not sufficient--- condition the true equation of motion
must fulfill. The true equation of motion, which
will not have neither preacceleration nor runaway solutions, is of second order and 
can only be constructed by using a series expansion or a method of successive approximations.

Others \cite{Villarroel97} consider that the commented difficulties with
Lorentz Dirac equation are not real physical problems, as they accept
that acceleration can have a singularity in points where the applied
force has a discontinuity. 
%
%

None of these justifications is fully satisfactory to us. Consider a
classical charge modelled by a charge distribution and the corresponding
energy-momentum distribution inside a sphere of radius $\epsilon$.
Provided that a suitable set of constitutive relations is added, the
local conservation of energy-momentum  yields an evolution law for this
continuous medium, which is deterministic and causal: the electric
current and the energy-momentum distribution at $t=0$ determine the
future values of these magnitudes. It is, to say the least, startling
that, on taking the limit $\epsilon\rightarrow 0$, the causal and
deterministic nature of the classical problem is lost.

Apparently Lorentz-Dirac equation is an unavoidable and flawless
consequence of classical electrodynamics plus the local conservation of
total energy and momentum \cite{Dirac38,Rohrlich65,Teitel79}. 
However, as the electromagnetic field contribution to
the energy-momentum tensor is singular on the charge's worldline ---it
behaves as $\Theta^{\mu\nu}\approx\,$O$(r^{-4})$--- some creative
``tricks'' are necessary to appropriately handle such a singular
behavior in the energy-momentum balance.  In our opinion, in most
approaches to this problem  some additional assumption slips into the
reasoning through one of these ``tricks''. 

In this context, it is worth mentioning Rowe's work \cite{Rowe75,Rowe78}, 
where more elaborated mathematical tools, namely
regularization of generalized functions, are used to properly handle the
singularity in $\Theta^{\mu\nu}$ and obtain the Lorentz-Dirac equation. 
The use of generalized functions  (or
distributions) has also the advantage that no mass renormalization is
necessary.

We shall here use these same mathematical tools to review the derivation
of Lorentz-Dirac equation and see that, contrary to the common belief,
it is not a straight consequence of classical electrodynamics plus
energy-momentum conservation, but it includes an elementary extra
assumption. 

We shall here describe a point charge as a current distribution in an
extended material body in the limit where the radius
$\epsilon\rightarrow 0$. The total energy-momentum tensor results from
two contributions: the electromagnetic part, $\Theta^{\mu\nu}$, which is
associated to the field and pervades spacetime, and the material part,
$K^{\mu\nu}$, which we assume confined to a world-tube of radius
$\epsilon$ and accounts for kinetic energy and the stresses that balance
the electric repulsion among the parts of a neat total charge confined
in a small volume.

For $\epsilon>0$ both contributions $\Theta^{\mu\nu}$ and $K^{\mu\nu}$
are continuous functions and can be considered separately. But in the
limit  $\epsilon\rightarrow 0$, the electromagnetic part presents a
singularity O$(r^{-4})$ on the worldline. Therefore, in the limit
$\epsilon\rightarrow 0$ none of these two contributions can be properly
defined, even resorting to generalized functions. However, nothing
forbids the total energy-momentum tensor to converge to a generalized
function for $\epsilon\rightarrow 0$, which will likely include $\delta$
functions and its derivatives on the point charge worldline.

In our approach we do not need to assume that the involved functions are
analytic. Although Taylor expansions to some finite order are used,
these hold for functions that are smooth enough, without need of
analyticity \cite{Apostol}.

We shall examine what restrictions on the charge's motion follow from
local conservation of energy and momentum, and find that the result is
not Lorentz-Dirac equation but a somewhat less restrictive equation,
already derived by Honig and Szamosi \cite{Szamosi81} by extending Dirac's work. 
Then we shall see
that this equation admits solutions that are free of both
preacceleration or runaways.

\section{Statement of the problem \label{S2}}

\subsection{Notation \label{SS2.1}}

The retarded Liénard-Wiechert field of a point charge has an outstanding
role along the present paper. Therefore it will be helpful to use
retarded optical coordinates \cite{SYNGE} (as in ref.\ \cite{Rowe75})
based on a timelike worldline $\Gamma \equiv\{z^\mu(\tau)\}$ and an
orthonormal tetrad $\{e^\mu_{(\alpha)}\}_{\alpha =1,2,3,4}$, which is
Fermi-Walker transported along $\Gamma$,
\begin{equation}
\frac{d e^\mu_{(\alpha)}}{d\tau} = [v^\mu a_\nu-v_\nu
a^\mu]\,e^\nu_{(\alpha)} \,.
\label{b1}
\end{equation}
With a properly chosen initial tetrad, the latter evolution equation is
consistent with the conditions
\begin{equation}  
        e^\mu_{(\alpha)} e^\nu_{(\beta)} \eta_{\mu\nu} =
             \eta_{\alpha\beta} \,,\qquad  
        e^\mu_{(4)} = v^\mu = \dot{z}^\mu \qquad {\rm and} \qquad a^\mu = 
             \dot{v}^\mu\,, 
\label{b2}
\end{equation}
where a `dot' means \guillemotleft derivative with respect to
$\tau${\guillemotright} and $\eta_{\mu\nu}=(+++-)$.
Moreover, from now on we use units such that $c=1$.

For any point $x$ in spacetime, the equation 
\begin{equation} 
[x^\mu - z^\mu(\tau)] [x^\nu - z^\nu(\tau)] \eta_{\mu\nu} = 0 \,,
\label{b3}
\end{equation}
supplemented with $x^4>z^4(\tau)$, has always a unique solution, $\tau
=\tau(x)$, which defines a time coordinate for $x$.

The space coordinates are
\begin{equation} 
X^i = e^\mu_{(i)} \left(x_\mu - z_\mu[\tau(x)]\right)
\label{b4}
\end{equation}
and the inverse coordinate transformation then reads
\begin{equation}
x^\mu = z^\mu(\tau) + \rho v^\mu(\tau) +X^i e^\mu_{(i)}(\tau)\,,
\label{b5}
\end{equation}
where $\rho = \|\vec{X} \| = \sqrt{(X^1)^2+(X^2)^2+(X^3)^2}$.

The following relations and quantities, introduced in ref.\ \cite{Teitel79}, will be
useful hereon: 
\begin{equation} 
\left.  \begin{array}{ll}
      \rho = - [x_\mu - z_\mu(\tau)] v^\mu(\tau)\,, & \quad 
            k^\mu :=\displaystyle{\frac1\rho \,[x^\mu - z^\mu(\tau)]}\,, \\
      n^\mu := k^\mu - v^\mu \,,& n^\mu n_\mu = 1 \,, \qquad k_\mu v^\mu = -1\,,
        \end{array}    \right\}
\label{b6}
\end{equation}
\begin{equation}
\partial_\mu\rho = n_\mu + \rho (a^\alpha n_\alpha) k_\mu\,.
\label{b7}
\end{equation}
The unit space vector $n^\mu$ can be written as
$$ n^\mu = \frac{X^i}{\rho}\,e^\mu_{(i)} \equiv \hat{n}^i e^\mu_{(i)} \,.$$

Finally, the relationship between the volume elements in Lorentzian and in retarded
optical coordinates is
\begin{equation}
d^4 x = d\tau\,d^3\vec{X} = \rho^2 d\tau\,d\rho\, d^2\Omega(\hat{n})\,,
\label{b8}
\end{equation}
where $d^2\Omega(\hat{n})$ is the solid angle element.

\subsection{Some definitions and postulates \label{SS2.2}}

A point charge is described by a current density four-vector, $j^\mu$,
and an energy-momentum tensor, $t^{\mu\nu}$, fulfilling
\begin{equation}
\partial_\mu j^\mu = 0 \, , \qquad \partial_\mu t^{\mu\nu} = 0 \qquad {\rm and} 
\qquad t^{\mu\nu} = t^{\nu\mu} \,,
\label{C2.1}
\end{equation}
respectively, the local conservation laws for total electric charge,
energy-momentum, and angular momentum.

We expect to obtain $j^\mu$ and $t^{\mu\nu}$  as the limit of continuous
distributions of charge and energy-momentum when the radius goes to
zero, namely, 
\begin{list}
{(\alph{llista})}{\usecounter{llista}}
\item{an electric current vector $J^\mu(\epsilon;x)$, which is confined to an
   ``optical tube'' of radius $\epsilon$ around a timelike worldline
    $\Gamma$, that is,  
        \begin{equation}
        \rho(x) > \epsilon \Rightarrow J^\mu(\epsilon;x) = 0 \,,
        \label{C2.0}
        \end{equation}
      where $\rho(x)$ is given by (\ref{b6}),}
\item{an energy-momentum tensor $T^{\mu\nu}(\epsilon;x)$ which results from two
      contributions: 
        \begin{equation}
           T^{\mu\nu}(\epsilon;x) = \Theta^{\mu\nu}(\epsilon;x) + K^{\mu\nu}(\epsilon;x)\,.
        \label{C2.2}
        \end{equation}
       The first term comes from the total electromagnetic field:
        \begin{equation}
           F^{\mu\nu}(\epsilon;x) = F_R^{\mu\nu}(\epsilon;x) + F_{ext}^{\mu\nu}(x)\,,
        \label{C2.2p}
        \end{equation}
       namely, the sum of the retarded solution of the Maxwell equations
       for the current $J^\mu(\epsilon;x)$ plus an external free electromagnetic
       field. The second term in (\ref{C2.2}) comes from the matter distribution
       which is also confined to the above mentioned ``optical tube'': 
        \begin{equation}
        \rho(x) > \epsilon \Rightarrow K^{\mu\nu}(\epsilon;x) = 0 \,.
        \label{C2.01}
        \end{equation}}
\end{list}
The above continuous distributions of electric current and
       energy-momentum are assumed to fulfill the local conservation laws: 
        \begin{equation}
        \partial_\mu J^\mu = 0 \, , \qquad \partial_\mu T^{\mu\nu} = 0 \,, 
        \qquad T^{\mu\nu} = T^{\nu\mu} \,.
        \label{C2.1p}
        \end{equation}
We shall assume that both $J^\mu(\epsilon;x)$ and $K^{\mu\nu}(\epsilon;x)$ are
locally summable in $\mathbb{R}^4$ and that $F_{ext}^{\mu\nu}(x)$ is continuous in
$\mathbb{R}^4$. 

The retarded electromagnetic field is given by \cite{Rohrlich65p}
\begin{equation}
F_R^{\mu\nu}(\epsilon;x) = \frac{8\pi}{c} \,\int \, J^{[\nu}(\epsilon;x)
\partial^{\mu]}D_R(x-x^\prime) \,d^4x^\prime  
\label{C2.3}
\end{equation}
with 
$$D_R(x)=\displaystyle{\frac1{2\pi}\,Y (x^4)\delta(x^\rho x_\rho)}
$$
[$Y(x^4)$ is the Heaviside step function.] The retarded electromagnetic
field is thus a continuous function and therefore locally summable in
$\mathbb{R}^4$. 

In its turn, the electromagnetic contribution to the energy-momentum tensor,
\begin{equation}
\Theta^{\mu\nu}(\epsilon;x) = \frac1{4\pi} \,\left[
F^{\mu\alpha}(\epsilon;x)\,F^\nu_{\;\alpha}(\epsilon;x) - \frac14 \eta^{\mu\nu}
F^{\rho\alpha}(\epsilon;x)\,F_{\rho\alpha}(\epsilon;x) \right] \,,
\label{C2.4}
\end{equation}
is also locally summable.

The framework where the limits for $\epsilon\rightarrow 0$ of
$J^\mu(\epsilon;x)$ and $T^{\mu\nu}(\epsilon;x)$ are mathematically meaningful and
can be appropriately handled is the space ${\cal D}^\prime(\mathbb{R}^4)$ of generalized
functions \cite{VLAD1,GELF}. As locally summable functions,
$J^\mu(\epsilon;x)$ and $T^{\mu\nu}(\epsilon;x)$ can be associated to generalized
functions and, provided that the limits 
$$ j^\mu = \lim_{\epsilon\rightarrow 0} J^\mu(\epsilon) \in {\cal D}^\prime(\mathbb{R}^4) \,, 
\qquad 
t^{\mu\nu} = \lim_{\epsilon\rightarrow 0} T^{\mu\nu}(\epsilon) \in {\cal D}^\prime(\mathbb{R}^4) $$
exist, the continuity of differentiation operators in ${\cal D}^\prime(\mathbb{R}^4)$ \cite{VLAD2} guarantees the conservation laws (\ref{C2.1}) as the limit of (\ref{C2.1p}) for $\epsilon\rightarrow 0$. 

These conservation laws must now be understood in the sense of ${\cal D}^\prime(\mathbb{R}^4)$,
i.\ e.\ $\forall \varphi \in {\cal D}(\mathbb{R}^4) \,,$
$$ (\partial_\mu j^\mu,\varphi) = 0 \qquad {\rm and} \qquad 
(\partial_\mu t^{\mu\nu},\varphi) = 0 $$
or 
\begin{equation}
(j^\mu,\partial_\mu \varphi) = 0 \qquad \mbox{and} \qquad (t^{\mu\nu},\partial_\mu \varphi) = 0 \,.
\label{C2.5}
\end{equation}

\section{The point charge limit \label{S3}}

\subsection{The electric current \label{SS3.1}}

If the support of $J^\mu(\epsilon;x)$ is the ``optical tube'' $\rho(x)\leq
\epsilon$, then for any $\varphi \in{\cal D}(\mathbb{R}^4)$ such that ${\rm supp}\,\varphi$ does
not intersect the worldline $\Gamma$, it exists $\epsilon_1>0$ such that
$\varphi(x) = 0$ whenever $\rho(x)\leq \epsilon_1$. Therefore, for all
$\epsilon<\epsilon_1$, 
$$\left(J^\mu(\epsilon),\varphi\right) = \int d^4x\, J^\mu(\epsilon;x)\,\varphi(x) = 0 \,,$$ 
and in the limit $\epsilon\rightarrow 0$ it follows that
$$(j^\mu,\varphi)=0\,,\quad \forall\varphi\in {\cal D}(\mathbb{R}^4)\qquad \mbox{such that}\quad \Gamma \cap{\rm supp}\,\varphi = \emptyset\,. $$ 

The support of the generalized function $j^\mu$ is therefore confined to
the worldline $\Gamma$ and, according to a well known result on
generalized functions \cite{VLAD3}, $j^\mu$ can be written as a sum of
$\delta$-functions and its derivatives up to a finite order: 
\begin{eqnarray}
j^\mu &=& \int d\tau\,\left[l^\mu(\tau)\,\delta(x-z(\tau)) + l^{\alpha\mu}(\tau)
      \,\partial_\alpha \delta(x-z(\tau)) + \ldots \right.  \nonumber \\
      & & \left. + l^{\alpha_1\ldots\alpha_n\mu}
(\tau)\,\partial_{\alpha_1\ldots\alpha_n}\delta(x-z(\tau)) \right] 
\label{C3.5p}
\end{eqnarray}
with $l^{(\alpha_1\ldots\alpha_r)\mu}\,v_{\alpha_1} = 0$\,;
\,$r=1,\ldots n$.

To model a point charge we only keep the lowest order term and, as a
consequence of the conservation law (\ref{C2.1}), we have \cite{Teitel79}
\begin{equation}
j^\mu = e\,\int d\tau\,v^\mu(\tau)\,\delta(x-z(\tau)) \,,
\label{C3.6}
\end{equation}
where $e$ is the electric charge of the particle and is a constant scalar.

\subsection{The energy-momentum tensor \label{SS3.2}}
In our approach, the limits for $K^{\mu\nu}(\epsilon)$ and $\Theta^{\mu\nu}(\epsilon)$ do not need to
exist separately  in ${\cal D}^\prime(\mathbb{R}^4)$. Our assumption is weaker and only the joint limit is assumed to be physically meaningful: 
\begin{equation}
t^{\mu\nu} = \lim_{\epsilon\rightarrow 0}\left[K^{\mu\nu}(\epsilon) + 
  \Theta^{\mu\nu}(\epsilon) \right] \in {\cal D}^\prime(\mathbb{R}^4) \,.
\label{C3.7}
\end{equation}
This fact expresses the notion that, although in the separate limits for both $K^{\mu\nu}(\epsilon)$ and $\Theta^{\mu\nu}(\epsilon)$ some infinities on the worldline $\Gamma$ could arise, these infinities will cancel each other, so that $t^{\mu\nu}$ is defined in ${\cal D}^\prime(\mathbb{R}^4)$.

\subsubsection{The matter contribution}
If we restrict to test functions $\varphi \in {\cal D}(\mathbb{R}^4-\Gamma)$, we have that
\begin{equation}
\lim_{\epsilon\rightarrow 0}\,K^{\mu\nu}(\epsilon) =0  \in {\cal D}^\prime(\mathbb{R}^4-\Gamma) \,.
\label{C3.7p}
\end{equation}
Indeed, for any $\varphi \in{\cal D}(\mathbb{R}^4-\Gamma)$ it exists $\epsilon_1>0$ such that
$\varphi(x)=0$ whenever $\rho(x) \leq \epsilon_1$. The confinement condition
(\ref{C2.01}) then implies that 
$$ \forall \epsilon<\epsilon_1\,,\quad (K^{\mu\nu}(\epsilon),\varphi)=\int d^4x\,K^{\mu\nu}(\epsilon;x) \,\varphi(x) =0   $$
and equation (\ref{C3.7p}) follows \cite{VLAD4}. 

\subsubsection{The electromagnetic contribution \label{SS3.3}} 
Recall now equations (\ref{C2.3}) and (\ref{C2.4}). We have the pointwise limit
\begin{equation}
\lim_{\epsilon\rightarrow 0}\,F^{\mu\nu}(\epsilon;x) = F_R^{\mu\nu}(x) +F_{\rm ext}^{\mu\nu}(x) \,,
\label{C3.8}
\end{equation}
where $F_R^{\mu\nu}(x)$ is the retarded Liénard-Wiechert field, and is
defined whenever $x\notin\Gamma$. It can be written as the sum of the
radiation field plus the velocity field: 
\begin{equation} 
F_R^{\mu\nu}(x) = F_I^{\mu\nu}(x) + F_{II}^{\mu\nu}(x) \,,
\label{C3.9}
\end{equation}
where, in the notation introduced in subsection \ref{SS2.1} (also in ref.\ \cite{Teitel79}):  
\begin{eqnarray}
F_I^{\mu\nu}(x) & = & \frac{2e}{\rho}\left[ (ak)\,v^{[\mu}k^{\nu]} + a^{[\mu}k^{\nu]}\right] \,,
\label{C3.10} \\
F_{II}^{\mu\nu}(x) & = & \frac{2e}{\rho^2}\,v^{[\mu}k^{\nu]} \,.
\label{C3.11}
\end{eqnarray}
(Here $(ak)\equiv a^\lambda k_\lambda$.)
Similarly, for the electromagnetic energy-momentum tensor we have the pointwise convergence: 
$$ \lim_{\epsilon\rightarrow 0}\,\Theta^{\mu\nu}(\epsilon;x) =\Theta^{\mu\nu}(x)\,,  
$$
except at the points $x\in \Gamma$.

As a consequence of (\ref{C3.8}), $\Theta^{\mu\nu}(x)$ can be splitted as
\begin{equation}
\Theta^{\mu\nu}(x) = \Theta_R^{\mu\nu}(x) + \Theta_{\rm ext}^{\mu\nu}(x) + \Theta_{\rm mix}^{\mu\nu}(x) \,.
\label{C3.12}
\end{equation}
The first and second terms in the r. h. s. respectively result from
substituting $F_R^{\mu\nu}(x)$ and $F_{\rm ext}^{\mu\nu}(x)$ into the
quadratic expression (\ref{C2.4}), whereas 
$\Theta_{\rm mix}^{\mu\nu}(x)$ comes from the cross terms.

$\Theta_{\rm mix}^{\mu\nu}(x)$ and $\Theta_{\rm ext}^{\mu\nu}(x)$ are
locally summable in $\mathbb{R}^4$. This is obvious for $\Theta_{\rm
ext}^{\mu\nu}(x)$ because it is continuous everywhere. As for
$\Theta_{\rm mix}^{\mu\nu}(x)$, it is a sum of products of $F_{\rm
ext}^{\mu\nu}(x)$, which is continuous, and $F_{R}^{\mu\nu}(x)$, which
is also continuous except for a singularity of order $\rho^{-2}$ on
$\Gamma$ that is cancelled by the factor $\rho^2$ in the volume element
(\ref{b8}). Therefore, $\Theta_{\rm mix}^{\mu\nu}(x)$ is also locally
summable in $\mathbb{R}^4$. We shall respectively denote: 
\begin{equation}
\theta_{\rm ext}^{\mu\nu} := \lim_{\epsilon\rightarrow 0}\,\Theta_{\rm ext}^{\mu\nu}(\epsilon;x)
\qquad{\rm and} \qquad
\theta_{\rm mix}^{\mu\nu}:=  \lim_{\epsilon\rightarrow 0}\,\Theta_{\rm mix}^{\mu\nu}(\epsilon;x)
\label{C3.12a} 
\end{equation}
with $\theta_{\rm ext}, \;\theta_{\rm mix}\in {\cal D}^\prime(\mathbb{R}^4)$.

Let us now consider the $\Theta_R^{\mu\nu}(x)$ contribution. It can be written as \cite{Teitel79} 
\begin{eqnarray}
\Theta_R^{\mu\nu}(x) &=& \frac{e^2}{4\pi\rho^4} \,\left[v^\mu k^\nu + v^\nu k^\mu + \frac12 \eta^{\mu\nu} - k^\mu k^\nu \right] + \nonumber \\
  & & \frac{e^2}{4\pi\rho^3} \,\left[a^\mu k^\nu + a^\nu k^\mu - (an) \left( 
       n^\mu k^\nu + n^\nu k^\mu \right) \right] + \nonumber \\
  & & \frac{e^2}{4\pi\rho^2} \left[a^2 -(an)^2\right] \,k^\mu k^\nu \,,
\label{C3.13}
\end{eqnarray}
which is continuous for $x\notin \Gamma$. 

Owing to the $\rho^{-4}$ and $\rho^{-3}$ singularities on the r.h.s.
of the above expression,
not only $\Theta_R^{\mu\nu}(x)$ has a singularity on $\Gamma$, but in
addition it is not locally summable. Therefore, no generalized function
in ${\cal D}^\prime(\mathbb{R}^4)$ can be associated to $\Theta_R^{\mu\nu}(x)$ in the
standard way. 

Now, since $\Theta_R^{\mu\nu}(x)$ is a continuous function on
$\mathbb{R}^4-\Gamma$, it is locally summable there, and this allows to
take its  finite part $\theta_R^{\mu\nu} \in {\cal D}^\prime(\mathbb{R}^4)$
\cite{VLAD5,GELF2}: 
\begin{equation}
(\theta_R^{\mu\nu},\varphi) \equiv \int d^4x\,\Theta_R^{\mu\nu}(x) \left[\varphi(x) - Y(L-\rho)\left[\varphi(z)+\rho k^\alpha\partial_\alpha\varphi(z)\right]\right] 
\label{C3.14}
\end{equation}
for any $\varphi\in {\cal D}^\prime(\mathbb{R}^4)$, where $L$ is an arbitrary chosen length scale, $z=z(\tau(x))$ and $\tau(x)$, $k^\alpha$ and $\rho(x)$ are defined in (\ref{b6}). 

Some points concerning the definition (\ref{C3.14}) are worth to comment:
\begin{list}
{(\roman{llista})}{\usecounter{llista}}
\item{The integral in the r.h.s. converges. Indeed, on the one hand, for
$\rho>L$, $\Theta_R^{\mu\nu}(x)$ is continuous and $\varphi(x)$ has
compact support and, on the other, inside $\rho\leq L$ we can use the
mean value Taylor theorem \cite{Apostol} for the smooth function
$\varphi$: 
$$ \varphi(x) =  \varphi(z) + \rho k^\lambda \partial_\lambda  \varphi(z) + \frac12 \rho^2 k^\lambda k^\mu \partial_{\lambda\mu}  \varphi(z+\rho^\prime k) $$ 
with $0<\rho^\prime<\rho(x)$. Now, since $\varphi$ is smooth and has
compact support, $\partial_{\lambda\mu}  \varphi$ is bounded and it
exists $M>0$ such that 
$$ \left| \varphi(x) - [ \varphi(z) + \rho k^\lambda \partial_\lambda \varphi(z)]\right| < M\rho^2 \,, \qquad 
x \in {\rm supp}\,\varphi\,, \quad 0\leq \rho \leq L\,.
$$ 
Hence the integrand in the r.h.s. of (\ref{C3.14}) presents a
singularity of order $\rho^{-2}$ on $\Gamma$ and therefore the integral
converges.} 
\item{For a test function $\varphi\in{\cal D}(\mathbb{R}^4-\Gamma)$, the function and all its derivatives vanish on
$\Gamma$. Hence, (\ref{C3.14}) amounts to
\begin{equation} 
(\theta_R^{\mu\nu},\varphi) = \int d^4x\,\Theta_R^{\mu\nu}(x) \,\varphi(x)\,.
< +\infty 
\label{C3.14p}
\end{equation} }
\item{The definition (\ref{C3.14}) consists of eliminating from the
integrand as many terms in the Taylor expansion of $\varphi(x)$ as necessary, in such a way that the remainder is summable and the condition
(ii) above is fulfilled. As a consequence, the finite part
$\theta_R^{\mu\nu} \in {\cal D}^\prime(\mathbb{R}^4)$ is not unique. Indeed, on the one hand, we
could have substracted some more terms in the Taylor expansion of
$\varphi$, and obtained a convergent integral also fulfilling the
requierement (ii). Besides, the length scale $L$ is quite arbitrary and
could even depend on $\tau(x)$. 

This results in that $\theta_R^{\mu\nu}$ is determined up to a finite
sum of $\Gamma$-supported $\delta$-functions and their derivatives,
multiplied by arbitrary $\tau$-dependent coefficients, in an expression
similar to (\ref{C3.5p}). We shall see that this lack of uniqueness in
the definition of $\theta_R^{\mu\nu}$ is not relevant at all, because we
are not actually interested in $\theta_R^{\mu\nu}$ but in the total
energy-momentum $t^{\mu\nu}$. Here lies the difference between our approach and that of Rowe \cite{Rowe78}.}
\end{list}

To give a more specific expression for $\theta_R^{\mu\nu}$, we realise
that since the r.h.s. of (\ref{C3.14}) is convergent, we can write
\begin{eqnarray*}
(\theta_R^{\mu\nu},\varphi) &=& \lim_{\epsilon\rightarrow 0} \left(\int
d^4x\,Y(\rho - \epsilon) \,\Theta_R^{\mu\nu}(x) \varphi(x) \right. \\
 & & \left. - \int
d^4x\,Y(\rho-\epsilon)\,Y(L-\rho)\, \Theta_R^{\mu\nu}(x)\left[\varphi(z)+\rho
k^\alpha\partial_\alpha\varphi(z)\right] \right)\,, 
\end{eqnarray*}
which after a short calculation leads to
\begin{equation}
\theta_R^{\mu\nu} = \hat\theta_R^{\mu\nu} - \int d\tau
\left([V^{\mu\nu} - \dot{U}^{\mu\nu}]\,\delta(x-z(\tau)) -
U^{\lambda\mu\nu}\,\partial_\lambda \delta(x-z(\tau))\right) \,,
\label{C3.15}
\end{equation}
where
\begin{equation}
\hat\theta_R^{\mu\nu} = \lim_{\epsilon\rightarrow 0} \left[
\Theta_R^{\mu\nu}(x) Y(\rho-\epsilon) - \frac{e^2}{\epsilon} \int d\tau
\left(\frac12 v^\mu v^\nu + \frac16 \hat\eta^{\mu\nu} \right)
\,\delta(x-z) \right] 
\label{C3.19}
\end{equation}
and the coefficients $V^{\mu\nu}$, ${U}^{\mu\nu}$ and $U^{\lambda\mu\nu}$ depend on $\tau$ and are:
\begin{eqnarray}
V^{\mu\nu}& = & -\frac{e^2}{6L} \left(3 v^\mu v^\nu + \hat\eta^{\mu\nu} \right) + \frac{2 e^2 L}{15} \left(5 a^2 v^\mu v^\nu + 2 a^2 \hat\eta^{\mu\nu} -  a^\mu a^\nu \right)  \,,
\label{C3.16} \\
U^{\mu\nu}& = & \frac23 e^2 L \left( a^\mu v^\nu +  a^\nu v^\mu \right) +\frac{e^2L^2}{15} \left( 5 a^2 v^\mu v^\nu + 2 a^2 \hat\eta^{\mu\nu} - a^\mu a^\nu \right)  \,,
\label{C3.17} \\
U^{\lambda\mu\nu} & = & \frac{e^2L}{15} \left( 3 a^\mu
\hat\eta^{\lambda\nu} + 3 a^\nu \hat\eta^{\lambda\mu} - 2 a^\lambda
\hat\eta^{\mu\nu} \right) + \nonumber \\ 
 & & \frac{e^2L^2}{15} \left( 2 a^2[v^\mu \hat\eta^{\lambda\nu} + v^\nu
\hat\eta^{\lambda\mu}] - a^\lambda[a^\mu v^\nu +  a^\nu v^\mu ]\right) \,.
\label{C3.18}
\end{eqnarray}
Notice that they depend on the length scale $L$. 

We shall hereafter write 
\begin{equation}
 \theta^{\mu\nu} = \theta_R^{\mu\nu} + \theta_{\rm ext}^{\mu\nu} +    \theta_{\rm mix}^{\mu\nu} \,.
\label{C3.20}
\end{equation}
Notice that $\theta^{\mu\nu}\in{\cal D}^\prime(\mathbb{R}^4)\subset {\cal D}^\prime(\mathbb{R}^4-\Gamma)$ . Now, since $\Theta^{\mu\nu}(x)$ is  locally summable in $\mathbb{R}^4-\Gamma$, it can be considered as a generalized function $\Theta^{\mu\nu}\in{\cal D}^\prime(\mathbb{R}^4-\Gamma)$ and, as a consequence of (\ref{C3.14p}) we have that
$$ \theta^{\mu\nu}=\Theta^{\mu\nu} \qquad {\rm in} \qquad  {\cal D}^\prime(\mathbb{R}^4-\Gamma) \,.$$ 

\subsubsection{The total energy-momentum tensor \label{SS3.4}}

The total energy-momentum tensor $t^{\mu\nu}$ is defined by the limit
(\ref{C3.7}). For any test function $\varphi\in{\cal D}(\mathbb{R}^4-\Gamma)$ we have, as a
consequence of (\ref{C3.7p}), that 
$$ (t^{\mu\nu},\varphi) = \lim_{\epsilon\rightarrow 0} \int d^4 x\; \Theta^{\mu\nu}(\epsilon,x)\,\varphi(x) $$ 
and, using (\ref{C3.12}), (\ref{C3.14p}) and (\ref{C3.20}), we obtain
$$ (t^{\mu\nu},\varphi) = (\theta^{\mu\nu},\varphi) \,,\qquad \forall \varphi\in{\cal D}(\mathbb{R}^4-\Gamma)\,. $$
Therefore, $t^{\mu\nu} -\theta^{\mu\nu} \in {\cal D}^\prime(\mathbb{R}^4)$ has support on $\Gamma$ and, according to a well known result \cite{VLAD3}, it can be written as a finite sum:  
\begin{eqnarray}
t^{\mu\nu} - \theta^{\mu\nu} &=& \int d\tau\,\left[m^{\mu\nu}(\tau)\,\delta(x-z(\tau)) +
m^{\alpha\mu\nu}(\tau)\,\partial_\alpha \delta(x-z(\tau)) + \ldots \right.  \nonumber \\
 & & \left. + m^{\alpha_1\ldots\alpha_n\mu\nu} (\tau)\, \partial_{\alpha_1\ldots\alpha_n} \delta(x-z(\tau)) \right]  \,,
\label{C4.21}
\end{eqnarray}
where 
$$ m^{(\alpha_1\ldots\alpha_r)\mu\nu} v_{\alpha_1} = 0 \,, \qquad r= 1 \ldots n \,.$$

So far there is no correspondence between $t^{\mu\nu} - \theta^{\mu\nu}$
and the, so to speak, ``matter contribution'' to the energy and
momentum. Therefore, we are not obliged to assign this difference the
value $m_0\int d\tau \,v^\mu v^\nu \,\delta(x-z(\tau)) $, as it is done
in ref.\ \cite{Rowe75,Rowe78}. However, for the sake of the
``elementarity'' of the point charge we shall retain as few terms in
(\ref{C4.21}) as possible, namely,  
$$ t^{\mu\nu} = \theta^{\mu\nu} + \int d\tau\,\left[m^{\mu\nu}(\tau)\,\delta(x-z(\tau)) +
m^{\lambda\mu\nu}(\tau)\,\partial_\lambda \delta(x-z(\tau)) \right]\,, $$
which combined with (\ref{C3.15}) and (\ref{C3.20}) leads to
\begin{equation} 
t^{\mu\nu} = \hat\theta_R^{\mu\nu} + \theta_{\rm ext}^{\mu\nu} + \theta_{\rm mix}^{\mu\nu} + t_s^{\mu\nu}  \,,
\label{C4.22}
\end{equation}
with 
\begin{equation}
t_s^{\mu\nu} \equiv \int
d\tau\,\left[p^{\mu\nu}(\tau)\,\delta(x-z(\tau)) +
p^{\lambda\mu\nu}(\tau)\,\partial_\lambda \delta(x-z(\tau)) \right] 
\label{C4.23}
\end{equation}
and 
$$ p^{\mu\nu} = m^{\mu\nu} + \dot{U}^{\mu\nu} - V^{\mu\nu} \,, \qquad
p^{\lambda\mu\nu} = m^{\lambda\mu\nu} + U^{\lambda\mu\nu} \,, $$
where, $p^{\lambda\mu\nu} v_\lambda = 0$ as it obviously follows from
(\ref{C3.18}) and (\ref{C4.21}). 

\section{Conservation laws and equations of motion \label{S5}}
The local conservation laws (\ref{C2.1}) will then yield some
restrictions on the coefficients $p^{\mu\nu}$ and $p^{\lambda\mu\nu}$
\cite{Mathis40,Havas76}. First of all, the symmetry of
$t^{\mu\nu}$ implies that 
$$ p^{\mu\nu} = p^{\nu\mu} \,, \qquad    p^{\lambda\mu\nu} =
p^{\lambda\nu\mu} \,. $$ 
Now, it is helpful to separate these coefficients in their components
respectively parallel and orthogonal to the velocity $v^\mu$: 
\begin{equation}
\left.
\begin{array}{l} 
p^{\mu\nu} = M v^\mu v^\nu + p^\mu v^\nu + p^\nu v^\mu + p_\perp^{\mu\nu}\,, \\
p^{\lambda\mu\nu} = Q^\lambda v^\mu v^\nu + Q^{\lambda\mu} v^\nu + Q^{\lambda\nu} v^\mu +     Q^{\lambda\mu\nu} \,,
\end{array}
\right\}
\label{C5.24}
\end{equation}
where all tensors and vectors other than $v^\mu$ are orthogonal to the
velocity. The local conservation law (\ref{C2.1}) then implies that
\begin{equation}
\partial_\mu \hat\theta_R^{\mu\nu} + \partial_\mu \theta_{\rm ext}^{\mu\nu} + \partial_\mu \theta_{\rm mix}^{\mu\nu} +\partial_\mu t_s^{\mu\nu} = 0\,.
\label{C5.25}
\end{equation}
Now, since $ \theta_{\rm ext}^{\mu\nu}$ is the  energy-momentum tensor
of a free electromagnetic field, $\partial_\mu  \theta_{\rm
ext}^{\mu\nu} = 0$. Similarly, the cross term contribution is  
\begin{equation}
\partial_\mu \theta_{\rm mix}^{\mu\nu} = - F_{\rm ext}^{\mu\nu} j_\mu = -e\, \int
d\tau\,F_{\rm ext}^{\mu\nu}(z)\, v_\mu(\tau) \,\delta(x-z(\tau)) \, 
\label{C5.26}
\end{equation}
and (see Appendix A [equation (\ref{A4})] for details)
\begin{equation}
\partial_\mu \hat\theta_{R}^{\mu\nu} = \frac23 e^2\, \int d\tau\,\left[a^2 v^\nu - 
\dot{a}^\nu\right]  \,\delta(x-z(\tau)) \,.
\label{C5.27}
\end{equation}

Finally, using (\ref{C5.24}) and after several integrations by parts, we also obtain 
\begin{eqnarray}
\partial_\mu t_s^{\mu\nu} 
 & = & \int d\tau \left[\frac{d\;}{d\tau} \left(M v^\nu + p^\nu +
a_\lambda [Q^\lambda v^\nu + Q^{\lambda\nu}]  \right)  \delta(x-z) \right.
\nonumber \\ 
 &+& \left.\left(v^\nu p^\mu + p_\perp^{\mu\nu} + \hat\eta^\mu_\lambda
\frac{d\;}{d\tau}[Q^\lambda v^\nu + Q^{\lambda\nu}] \right)
\partial_\mu\delta(x-z) \right.\nonumber \\
& + & \left.  \left(Q^{\lambda\mu\nu} + Q^{\lambda\mu} v^\nu \right) \partial_{\lambda\mu}\delta(x-z) \right]  
\label{C5.28} 
\end{eqnarray}
and, substituting (\ref{C5.26}), (\ref{C5.27}) and (\ref{C5.28}) into (\ref{C5.25}), we arrive at 
\begin{eqnarray}
0 & = & \int d\tau \left[\left\{\frac{d\;}{d\tau} \left(M v^\nu + p^\nu + a_\lambda [Q^\lambda v^\nu + Q^{\lambda\nu}] \right) \right. \right. \nonumber \\
& + & \left. \left. \frac23 e^2 (a^2 v^\nu -\dot a^\nu) - F^\nu  \right\} \delta(x-z)  \right. \nonumber \\ 
 &+& \left.\left(v^\nu p^\mu +p_\perp^{\mu\nu} + \hat\eta^\mu_\lambda
\frac{d\;}{d\tau}[Q^\lambda v^\nu + Q^{\lambda\nu}] \right)
\partial_\mu\delta(x-z) \right. \nonumber \\
 & + & \left. \left(Q^{\lambda\mu\nu} + Q^{\lambda\mu} v^\nu \right)\partial_{\lambda\mu}\delta(x-z) \right]   \,,
\label{C5.29} 
\end{eqnarray}
where $F^\nu \equiv e F_{\rm ext}^{\mu\nu}(z) v_\mu$.

As the derivatives of $\delta$-functions in the r.h.s. are contracted
with tensors that are transversal to the worldline, each term must
vanish separately and therefore
\begin{eqnarray}
\frac{d\;}{d\tau} \left(M v^\nu + p^\nu + a_\lambda [Q^\lambda v^\nu +
Q^{\lambda\nu}] \right)+ \frac23 e^2 (a^2 v^\nu -\dot a^\nu) &=& F^\nu\,,
\label{C5.30} \\ 
v^\nu p^\mu +p_\perp^{\mu\nu} + \hat\eta^\mu_\lambda
\frac{d\;}{d\tau}[Q^\lambda v^\nu + Q^{\lambda\nu}] &=& 0\,,    \label{C5.31}
\\ 
Q^{(\lambda\mu)\nu} + Q^{(\lambda\mu)} v^\nu & = & 0 \,.\label{C5.32} 
\end{eqnarray}

Since $Q^{\lambda\mu}$ and $Q^{\lambda\mu\nu}$ are orthogonal to
$v_\lambda$ and  $Q^{\lambda\mu\nu}=Q^{\lambda\nu\mu}$, equation
(\ref{C5.32}) implies that  
\begin{equation}
Q^{(\lambda\mu)} = 0 \qquad {\rm and} \qquad    Q^{\lambda\mu\nu} = 0\,.
\label{C5.33}
\end{equation}
Substituting this into (\ref{C5.31}), we obtain
\begin{eqnarray}
p^\mu &=&- \dot Q^\mu + v^\mu Q^\lambda a_\lambda - Q^{\mu\lambda} a_\lambda \, , \label{C5.34} \\
p_\perp^{\mu\nu} &=&- Q^\mu a^\nu -\dot Q^{\mu\nu} + v^\nu Q^{\mu\lambda} a_\lambda + v^\mu Q^{\lambda\nu} a_\lambda\,.
\label{C5.34a}
\end{eqnarray}
Since $p_\perp^{\mu\nu}$ is symmetric and $Q^{\mu\nu}$ is skewsymmetric, it follows that
\begin{equation}
p_\perp^{(\mu\nu)} = - Q^{(\mu} a^{\nu)} 
\label{C5.35a}
\end{equation}
and
\begin{equation}  
\dot Q^{\mu\nu} = -Q^{[\mu} a^{\nu]}- 2 v^{[\mu} Q^{\nu]\lambda} a_\lambda \,.
\label{C5.35}
\end{equation}
Finally, substituting (\ref{C5.33}), (\ref{C5.34}) and (\ref{C5.35})
into (\ref{C5.30}),  after a short manipulation we arrive at
\begin{equation} 
\frac{d\;}{d\tau} \left(\left[M + 2 Q^\lambda a_\lambda\right] v^\nu - \dot{Q}^\nu + 2 Q^{\lambda\nu} a_\lambda \right)+  
\frac23 e^2 (a^2 v^\nu -\dot a^\nu) = F^\nu  \,.
\label{C5.36} 
\end{equation}

On the basis of solely  the conservations of energy-momentum and angular
momentum we have thus found that
\begin{list}
{(\alph{llista})}{\usecounter{llista}}
\item the quantities $M, \,Q^\lambda \ldots , Q^{\lambda\mu\nu}$ in
equations (\ref{C5.24}) can be written in terms of  only ten independent
particle variables: $M$, $Q^\lambda$ and $Q^{[\lambda\mu]}$, that,
\item together with the worldline variables $z^\mu(\tau)$,
$v^\mu(\tau)$, \ldots  are subject to the differential system
(\ref{C5.35})--(\ref{C5.36}).
\end{list}
 
\subsection{Total momentum and angular momentum \label{SS.TM}}
 
Next, to have a clue of the physical meaning of $M$, $Q^\lambda$ and
$Q^{\lambda\nu}$, we examine the total linear and angular momenta.

The total linear momentum contained in the hypersurface
$\Gamma\equiv\{\tau={\rm constant}\}$ in  the optical coordinates
(\ref{b5}), i.\ e.\ the future light cone with vertex in $z^\mu(\tau)$,
is
\begin{equation} \label{C6.1}
P^\mu(\tau) = \int_\Gamma d\Sigma_\nu\, t^{\mu\nu} \qquad{\rm with} \qquad d\Sigma_\nu = -k_\nu \,d^3\vec{X}\,.
\end{equation}
Including now (\ref{C4.22}), we have that the total momentum $P^\mu$
results from three contributions:
$$ P^\mu = P_{\rm p}^\mu + P_{\rm mix}^\mu + P_{\rm ext}^\mu \,, $$
where $P_{\rm mix}^\mu$ and $P_{\rm ext}^\mu$ respectively come from the cross term 
$\theta_{\rm mix}^{\mu\nu}$ and the external field term $\theta_{\rm
ext}^{\mu\nu}$ in the energy-momentum tensor, and 
\begin{equation} \label{C6.2}
 P_{\rm p}^\mu = - \int_\Gamma d^3\vec{X}\, k_\nu ( t_s^{\mu\nu} +\theta_R^{\mu\nu} )
\end{equation}
is the contribution from the charge, i.\ e.\ the charge and its inseparable self-field.

On substituting (\ref{C3.19}), (\ref{C3.13}), (\ref{C4.23}) and
(\ref{C5.24}) into (\ref{C6.2}), after a little calculation we obtain
\begin{equation} \label{C6.3}
 P_{\rm p}^\mu = M v^\mu -\dot{Q}^\mu + \frac43 a_\lambda (Q^\lambda v^\mu + Q^{\lambda \mu})\,.
\end{equation}

Similarly, the total angular momentum in the hypersurface $\Gamma$, 
$$J^{\mu\nu}(\tau) = -\int_\Gamma d^3\vec{X}\, k_\sigma\, \left(x^\mu t^{\nu\sigma} - x^\nu t^{\mu\sigma} \right) \,,$$
comes from three contributions as well: $J^{\mu\nu} = J_{\rm p}^{\mu\nu}
+ J_{\rm mix}^{\mu\nu} + J_{\rm ext}^{\mu\nu}$. A similar calculation
yields the point charge contribution 
$$ J_{\rm p}^{\mu\nu}= z^\mu P_{\rm p}^\nu-z^\nu P_{\rm p}^\mu +S_{\rm p}^{\mu\nu} \,, $$
where
\begin{equation} \label{C6.4}
S_{\rm p}^{\mu\nu} = -2 Q^{[\mu} v^{\nu]} - 2 Q^{[\mu\nu]} 
\end{equation}
is the particle internal angular momentum. The second term on the r. h.
s. is orthogonal to the velocity and is the spin of the particle. On its
turn, the possibility that $Q^\mu \equiv - v_\nu S_{\rm p}^{\mu\nu}\neq
0$  is related with the fact that the center of motion \cite{Havas76}
does not necessarily lies on the particle's worldline.

To model a spinless charge, we choose $Q^{\mu\nu}=0$. Equation (\ref{C5.35}) then yields
\begin{equation} \label{C6.5}
Q^\mu = Q a^\mu
\end{equation}
and (\ref{C5.36}) can be further simplified to:
\begin{equation} \label{C6.6}
\frac{d\;}{d\tau} \left(\left[M+ 2 Q^\lambda a_\lambda\right] v^\nu - \dot{Q}^\nu \right)+  
\frac23 e^2 (a^2 v^\nu -\dot a^\nu) = F^\nu  \,.
\end{equation}
This agrees with the equation obtained by Honig and
Szamosi: (\ref{C6.6}) is equation (7) in \cite{Szamosi81}, with
$m=M+2Q a^2 - \ddot Q$, $R= 2\dot Q$ and $S= Q$. Lorentz-Dirac equation
is a particular case for $Q=0$.

\subsection{Summary}
A classical spinless point charge is therefore described by
\begin{list}
{(\alph{llista})}{\usecounter{llista}}
\item the electric current density (\ref{C3.6}) 
$$ j^\mu = e\,\int d\tau\,v^\mu(\tau)\,\delta(x-z(\tau)) \,,$$
where the electric charge $e$ is a constant scalar, and 
\item the total energy-momentum tensor (\ref{C4.22})
$$ t^{\mu\nu} = t_s^{\mu\nu} + \hat\theta_R^{\mu\nu} + \theta_{\rm ext}^{\mu\nu} + \theta_{\rm mix}^{\mu\nu} $$   
where $\hat\theta_R^{\mu\nu}\,$ and $t_s^{\mu\nu}\,$ are respectively given by (\ref{C3.19}) and (\ref{C4.23}), with 
\begin{eqnarray} \label{C6.7a}
p^{\lambda\mu\nu} &=& Q^\lambda v^\mu v^\nu \,,\qquad \qquad\qquad Q^\lambda = Q a^\lambda\,,  \\
p^{\mu\nu} &=& (M+ 2 Q^\lambda a_\lambda)v^\mu v^\nu - 2 \frac{d\;}{d\tau}\,(Q^{(\mu} v^{\nu)}) + Q^\mu a^\nu \;. \label{C6.7b}
\end{eqnarray}
\end{list}
The scalar variables $M$ and $Q$, together with the worldline
$z^\mu(\tau)$ are subject to equation (\ref{C6.6}), which has been
derived on the only basis that linear and angular momenta are conserved,
supplemented with the point limit and the assumption that the particle
is spinless.

\section{The equation of motion \label{C8}}
Equation (\ref{C6.6}) does not yield the law of motion yet. Indeed, it
consists of four equations for five unknowns, namely, $M$, $Q$ and
$z^\mu$ with the constraint $v^\mu v_\mu = -1$. The motion of the
particle is therefore underdetermined. 

This should not be surprising. The problem in dynamics of continuous
media for $\epsilon >0$, as we have posed it, is itself underdetermined,
because no constitutive equation has been assumed for the material
sustaining the electric charge, contrary, for instance, to what is done 
in \cite{Yaghjian,Moniz74}, were it is assumed that the
charge is rigidly distributed over a spherical shell of radius $\epsilon$.

Instead of advancing a matter constitutive equation for $\epsilon >0$,
then reexamining the problem and taking the limit $\epsilon\rightarrow
0$ to determine a final equation of motion, we shall directly posit a
{\em constitutive relation} connecting $M$, $Q$ and the worldline
invariants (curvature, torsion, etc.). 

Notice that, although it is the simplest choice and looks suitable for
an elementary charge, a prescription like $Q=0$ is not an appropriate
constitutive relation. Indeed, with a choice like this, (\ref{C6.6})
becomes Lorentz-Dirac equation which leads to the dilemma of solutions
that are either preaccelerated or runaway.

We shall base our guess of a constitutive relation on the requirements that
\begin{list}
{(\alph{llista})}{\usecounter{llista}}
\item it connects $M$, $Q$, $a^\nu$ and maybe some of their derivatives,  
\item when $a^\nu$, $Q$ and also their derivatives vanish, then $M=m_0$, and
\item if the point charge is acted by an external force $ F^\nu $ that 
vanishes for $\tau<0$ and for $\tau > \tau_1$, then: 
  \begin{itemize}
  \item $a^\nu(\tau)=0$, $M(\tau)=m_0$ and $Q(\tau)=0$ for $\tau <0$ and 
  \item $a^\nu \rightarrow 0$, $M \rightarrow m_0$ and $Q a^\nu\rightarrow 0$ asymptotically in the future.
  \end{itemize}
\end{list}
(The proper mass has the same value $m_0$ in the infinite past and
future, because we are assuming that the particle ``identity'' is
finally preserved.)


\subsection{Rectilinear motion \label{S6}}
To see whether a constitutive relation can be prescribed so that
(\ref{C6.6}) admits solutions that are neither runaway nor
preaccelerated, we shall examine the case of rectilinear motion. (Recall
that even in this simple case Lorentz-Dirac equation is not
satisfactory.)

Consider a point charge that initially is unaccelerated and free. Then, during
the interval $0\leq \tau \leq \tau_1$, it is acted by an external force
in a constant direction along the $X^1$ axis. The charge worldline will
remain in the plane $X^1X^4$ in spacetime and therefore,
$$\frac{dv^\mu}{d\tau}= a \,\hat{a}^\mu \qquad {\rm and} \qquad 
  \frac{da^\mu}{d\tau}= \dot{a} \,\hat{a}^\mu + a^2 \,v^\mu \,,$$
where $\hat{a}^\mu$ is the unit vector parallel to $a^\mu$, i.\ e.\ the first normal to the worldline.  The coefficients $p^{\mu\nu}$  and $p^{\lambda\mu\nu}$ in equations (\ref{C6.7a}) and (\ref{C6.7b}), i.\ e.\ the particle's contribution to the energy-momentum tensor are
\begin{eqnarray} \label{C7.7a}
p^{\lambda\mu\nu} &=& q \hat{a}^\lambda v^\mu v^\nu \,,\qquad \qquad\qquad Q^\lambda = Q a^\lambda \,, \\
p^{\mu\nu} &=& M\,v^\mu v^\nu - \dot{q}\,(\hat{a}^\mu v^\nu +\hat{a}^\nu v^\mu) + qa\,\hat{a}^\mu \hat{a}^\nu \,, \label{C7.7b}
\end{eqnarray}
with $q\equiv Qa$.

In this case, the only non-vanishing components of equation (\ref{C6.6}) are
\begin{equation}  \label{C7.1}  \left.
\begin{array}{lcl}
(\parallel v^\mu) & \quad & \displaystyle{\frac{d\;}{d\tau}\left(M+qa\right) = a \dot q \,,}\\
(\perp v^\mu) & \quad &  \displaystyle{a\,\left(M+qa\right) -\ddot{q} - \frac23 e^2\dot a = F\,.}
\end{array} \right\}
\end{equation}

These two equations must be supplemented with a constitutive relation
$M=M(a,q,\dot q)$ in order that evolution is determined. The phase space
is therefore coordinated by $(a,q,\dot q)$.

We would  expect that while the charge is not acted by any force,
$F(\tau)=0$, $-\infty<\tau<0$, then it remains in a state of uniform
rectilinear motion and the energy-momentum tensor is the one
corresponding to a free particle together with its Coulomb field, i.\
e.\ equations (\ref{C4.22}), (\ref{C6.7a}) and (\ref{C6.7b}) with
\begin{equation}  \label{C7.2}
a(\tau)=0\,, \qquad M(\tau)=m_0\,, \qquad q(\tau)= \dot q(\tau)=0\,, \qquad -\infty<\tau<0
\end{equation}
If an external force is then switched on: $F(\tau)\neq 0$,
$0\leq\tau<\tau_1$, then $a$, $M$, $q$ and $\dot q$ evolve according to
(\ref{C7.1}) with the initial data inferred from (\ref{C7.2}) and the
continuity of the orbit in phase space. This determines 
\begin{equation}  \label{C7.2a}
a(\tau) \,, \qquad M(\tau) \,, \qquad q(\tau) \quad {\rm and} \quad \dot q(\tau) \quad {\rm for} \quad 0<\tau<\tau_1
\end{equation}
After that the particle is not acted by a force any more and what we
would expect is that it asymptotically tends towards a free state, i. e.
$$ a(\tau)\rightarrow 0\,, \qquad M(\tau)\rightarrow m_0\,, \qquad
q(\tau)\rightarrow 0 \,, \qquad \dot q(\tau)\rightarrow  0\qquad {\rm
for}\quad  \tau \rightarrow \infty 
$$
(with the same asymptotical value $m_0$ for the mass, in order that the
particle's ``identity''  is preserved).

A way to achieve this behaviour consists in that the dynamical system
(\ref{C7.1}) supplemented with the constitutive relation has only one
equilibrium point for $a=q=\dot q=0$, which is asymptotically stable and
$M(0,0,0)=m_0$.

\subsection{A dynamical system \label{SS6.1}}
Using the constant $\displaystyle{\tau_0 \equiv \frac{2e^2}{3 m_0}}$, 
we introduce the new dimensionless variables
\begin{equation}  \label{C7.0}
 t\equiv \frac{\tau}{\tau_0} \,, \qquad 1+\mu \equiv \frac{M+qa}{m_0} \,, \qquad 
\alpha \equiv a\,\tau_0  \,, \qquad 
\rho \equiv \frac{q}{m_0\tau_0} 
\end{equation}
and reduce (\ref{C7.1}) with $F=0$ to the simpler equivalent system
$$ \mu^\prime=  a \rho^\prime \,, \qquad
\rho^{\prime\prime}+\alpha^\prime = \alpha (1+\mu) \,, \qquad
\mu=\mu(\alpha,\rho,\rho^\prime) \,,
$$
where `prime' means \guillemotleft derivative with respect to
$t$\guillemotright.

Then, by differentiating the constitutive relation and introducing the
variable $x\equiv \rho^\prime + \alpha$, we obtain
\begin{equation}  \label{C7.3}  \left.
\begin{array}{l}
\rho^\prime = x-\alpha\,,  \\
x^\prime = \alpha( 1+\mu)\,, \\
\alpha^\prime = A(\alpha, \rho, x)\,,
\end{array} \right\}
\end{equation}
where $$A(\alpha, \rho, x)\equiv \displaystyle{\frac1\mu_\alpha\left[(x-\alpha)(\alpha - \mu_\rho) - \alpha \mu_x (1+\mu) \right]}.$$

This dynamical system is already in normal form and is defined in the
 entire phase space  provided that the function $A(\alpha, \rho, x)$ has
 no singularities. Particularly, if we choose $\mu$ so that is a
 solution of  
 \begin{equation} \label{C7.5} A_0(\alpha,\rho,x) \,\mu_\alpha +
 (x-\alpha) \,\mu_\rho + \alpha\,(1+\mu) \,\mu_x = \alpha (x-\alpha) 
 \end{equation}
 with $A_0(\alpha,\rho,x)=l\alpha + p\rho+r x$ ($l$, $p$ and $r$
 constant) and $\mu(0,0,0)=0$, then the dynamical system (\ref{C7.3})
 becomes
 \begin{equation}   \label{C7.6} \frac{d\;}{dt}\left(\begin{array}{c}
 \alpha \\ \rho \\ x  \end{array} \right) =  \left(\begin{array}{ccc} l
 & p & r \\ -1 & 0 & 1 \\ 1 & 0 & 0  \end{array} \right) \,
 \left(\begin{array}{c} \alpha \\ \rho \\ x  \end{array} \right)  +
 \left(\begin{array}{c} 0 \\ 0 \\ \mu \alpha  \end{array} \right) \,.
 \end{equation} 
 If
 $p\neq 0$, the equilibrium points are
 \begin{eqnarray}  
 P_I&: \qquad
 & \alpha=\rho = x =0\,,  \nonumber \\ P_{II}&: \qquad & x=\alpha=\alpha_0
 \,, \quad \rho_0= - \frac{l+r}{p} \alpha_0 \quad {\rm and} \quad
 \mu(\alpha_0, \rho_0, \alpha_0)=-1\,.  \nonumber 
\end{eqnarray}
 Moreover, the constants $l$, $p$ and $r$ can be chosen so that the
 characteristic equation at $P_I$,  
 $$ X^3 - l X^2 +(p-r) X - p =0 \,,$$
 has three negative solutions and hence $P_I$ is an asymptotically
 stable equilibrium point. 

In Appendix B [equation (\ref{e7})] we see how a solution
$\mu=\mu(\alpha,\rho,x)$ of equation (\ref{C7.5}) that vanishes at
$P_I=(0,0,0)$ can be perturbatively obtained and is valid at least in a
neigbourhood of this phase point. 

Now, (\ref{C7.0}) can be used to obtain the constitutive equation
\begin{equation} \label{C7.7}
M= m_0 - q a m_0\mu \left(a \tau_0, \frac{q}{m_0\tau_0}, a\tau_0 +\frac{\dot q}{m_0}\right) \,.
\end{equation}
This, together with equations (\ref{C7.1}), determines a motion of the
charge that is free of both preacceleration and runaways, provided that
the force $F$ acts only during a finite interval of time. Indeed, if the
charge is unaccelerated in past infinity it remains so until its state 
is altered because $F$ has started to act. Then, when the force ceases,
the charge tends  to the asymptotically stable equilibrium point $a=0$,
$q= \dot q =0$, at least if the system was close enough when the force dissapeared.  

\section{Conclusion}
By studying the energy-momentum balance of a classical point charge with
the electromagnetic field, we have obtained that
\begin{list} 
{(\alph{llista})}{\usecounter{llista}}
\item the total energy-momentum tensor consists of (i) a regular part,
which comes from the external field contribution plus the regularization
of the self-field contribution, and (ii)
a singular part, with support on the charge worldline.
\item This singular part depends on two scalar coefficients $M(\tau)$
and $Q(\tau)$ and on the worldline variables $v^\mu(\tau)$,
$a^\mu(\tau)$, \ldots
\item These variables are constrained to fulfill the Honig-Szamosi
equation \cite{Szamosi81}, i.\ e.\ (\ref{C6.6}).
\end{list}
Lorentz-Dirac equation is obtained only if the constitutive relation
$Q=0$ is set by hand. The well known troubles that suffers the
Lorentz-Dirac equation are due to this bad choice rather than to
energy-momentum conservation itself.

We have then seen that, at least in the case of rectilinear motion, it
is possible to find a constitutive relation $M=M(a,Q,\dot Q)$ which,
together with equation (\ref{C6.6}) yields an equation of motion for the
point charge that is free from both preacceleration and runaways. That
is, if a charge is initially at rest, with proper mass $m_0$, and is
acted by an external force which lasts only a finite interval of time,
then there is no acceleration before the force starts and, when its
action ceases, the motion tends asymptotically to be rectilinear uniform
and the proper mass tends to $m_0$.

\section*{Acknowledgment}
The work of J.Ll.\ and A.M.\ is supported by Ministerio de Ciencia y
Tecnolog\'{\i}a,  BFM2003-07076, and Generalitat de Catalunya,
2001SGR-00061 (DURSI).
J.M.A.\ was supported by the University
of the Basque Country, UPV00172.310-14456/2002, and Ministerio de Educación y
Ciencia, FIS2004-01626.

\section*{Appendix A: Detailed computation of Eq.\ (\ref{C5.27})}

Using the definition (\ref{C3.19}), we have that $\forall \varphi\in{\cal D}^\prime(\mathbb{R}^4)$
\begin{eqnarray} 
(\partial_\mu \hat\theta_{R}^{\mu\nu},\varphi) = -(\theta_{R}^{\mu\nu},\partial_\mu \varphi) &=& 
 \lim_{\epsilon\rightarrow 0}\left\{-\int_{\rho\geq\epsilon} d^4x\;\Theta_R^{\mu\nu}(x)\partial_\mu\varphi(x) \right. + \nonumber \\  \label{A1}
 & &\qquad \left. \frac{e^2}{2\epsilon} \int_{-\infty}^\infty d\tau\,\left[v^\mu v^\nu+\frac13\hat\eta^{\mu\nu}\right]\partial_\mu\varphi\right\}.\qquad
\end{eqnarray}

Since $\Theta_R^{\mu\nu}(x)$ is summable for $\rho\geq\epsilon$, the
first integral on the r.h.s. becomes
$$ I_1\equiv \int_{\rho\geq\epsilon}
d^4x\;\partial_\mu\Theta_R^{\mu\nu}(x)\varphi(x)
-\int_{\rho\geq\epsilon}
d^4x\;\partial_\mu\left[\Theta_R^{\mu\nu}(x)\varphi(x)\right]\,.
$$
The first term vanishes because there is no current in
$\rho\geq\epsilon$ and, applying Gauss theorem, the second one yields
\begin{equation}  \label{A2}
\epsilon^2\,\int_{-\infty}^\infty d\tau\,\int \,d^2\Omega
\,\Theta_R^{\mu\nu}(\rho=\epsilon)\,[n_\mu + \epsilon(an)
k_\mu]\,\varphi(z^\lambda+\epsilon k^\lambda)\,,
\end{equation}
where $(an)\equiv a^\lambda n_\lambda$ and $d^2\Omega$ is the solid angle element. 
Using then equation (\ref{C3.13}) and the Taylor expansion
\cite{Apostol} $\varphi(z+\epsilon k) = \varphi(z) +\epsilon
k^\lambda\partial_\lambda\varphi(z)+\frac12 \epsilon^2 k^\mu k^\lambda
\partial_{\mu\lambda}\varphi(z)+{\rm O}(\epsilon^3) $, equation
(\ref{A2}) yields
\begin{eqnarray}
I_1&=&\int_{-\infty}^\infty d\tau\,\int \,\frac{d^2\Omega}{4\pi}\,
\left\{-\frac{e^2}{2\epsilon^2}\Big( v^\nu \epsilon (an)
[\varphi+\epsilon k^\lambda\partial_\lambda\varphi]\right.
\nonumber \\
  &+& 
n^\nu [1+\epsilon(an)]\,[\varphi+\epsilon k^\lambda\partial_\lambda\varphi+\frac12\epsilon^2 k^\mu k^\lambda\partial_{\mu\lambda}\varphi] \Big)  \nonumber \\
&+& \left. \frac{e^2}{\epsilon}\,[a^\nu - (a^2) n^\nu]\,[\varphi+\epsilon k^\lambda\partial_\lambda\varphi] + e^2 \,[a^2 -(an)^2] k^\nu\,\varphi \right\} + {\rm O}(\epsilon)\,. \nonumber
\end{eqnarray}
On integration with respect to $d^2\Omega$ and using that 
$$ \int \,d^2\Omega\, n^\nu = \int \,d^2\Omega \,n^\nu n^\mu n^\lambda = 0   
\qquad {\rm and} \qquad\int \,d^2\Omega \,n^\nu n^\mu= \frac{4\pi}3 \hat\eta^{\nu\mu} \,,$$
we arrive at
\begin{equation}  \label{A3}
I_1 = \frac{e^2}{2\epsilon}\,\int_{-\infty}^\infty d\tau\,  \left(a^\nu \varphi -\frac13 \hat\eta^{\mu\nu}\partial_\mu\varphi \right) + \frac{2 e^2}{3}\,\int_{-\infty}^\infty d\tau\,[a^2 v^\nu -\dot a^\nu ] \varphi\,.
\end{equation}
It is straightforward to check that the first term on the r.h.s. exactly
compensates the second term on the r.h.s. in (\ref{A1}). Therefore we
have
\begin{equation} \label{A4}
\partial_\mu \hat\theta_{R}^{\mu\nu}= 
\frac23 e^2\, \int d\tau\,\left[a^2 v^\nu -  \dot{a}^\nu\right]  \,\delta(x-z(\tau)) \,. \end{equation}

\section*{Appendix B: The constitutive relation}
We have to solve equation (\ref{C7.5})
\begin{equation} \label{e5}
(l \alpha + p \rho + r x)\,\mu_\alpha + (x-\alpha) \,\mu_\rho + \alpha\,(1+\mu) \,\mu_x = \alpha (x-\alpha) 
\end{equation}
with the ``initial condition'' $\mu(0,0,0)=0$.

It is easily seen that this equation admits a perturbative solution like
$$ \mu = \sum_{n=1}^\infty \mu^{(n)} $$
$\mu^{(n)}$ being a polynomial in the variables $a,q,x$ which is
homogeneous and has degree $2n$. If we write
$$\hat D \equiv (l \alpha + p \rho + r x)\partial_\alpha +(x-\alpha)\partial_\rho + \alpha\partial_x $$
then equation (\ref{e5}) yields the hierarchy:
\begin{eqnarray} \label{e6a}
 & & \hat D \mu^{(1)} = \alpha (x-\alpha)\,, \\
\label{e6b}
 n>1 & \quad & \hat D \mu^{(n)} = - \sum_{s=1}^\infty \mu^{(n-s)} \, \alpha\,\partial_x \mu^{(s)}\,.
\end{eqnarray}

The lowest order is relatively easy to solve and yields:
\begin{equation} \label{e7}
\mu = - \frac1{2\Delta}\left[ (p-r) \alpha^2 + p^2 \rho^2 + (r^2+p+rl) x^2 - 2 p \alpha x + 2 r p \rho x \right]+{\rm O}(4).
\end{equation}

\end{document}